\documentclass[twocolumn,pra,aps,superscriptaddress]{revtex4}
\usepackage{graphicx, color, fancybox}
\usepackage{amsfonts}
\usepackage{amssymb,amsmath,amsthm}

\begin{document}
\title{Unitary-process discrimination with error margin}
\author{T.~Hashimoto}
\affiliation{Department of Applied Physics University of Fukui, Fukui 910-8507, Japan}
\author{A.~Hayashi} 
\affiliation{Department of Applied Physics University of Fukui, Fukui 910-8507, Japan}
\author{M.~Hayashi}
\affiliation{Graduate School of Information Sciences, 
Tohoku University, Aoba-ku, Sendai, 980-8579, Japan}
\affiliation{Centre for Quantum Technologies, National University of Singapore, 
3 Science Drive 2, Singapore 117542}
\author{M.~Horibe}
\affiliation{Department of Applied Physics University of Fukui, Fukui 910-8507, Japan}

\begin{abstract}
We investigate a discrimination scheme between unitary processes. 
By introducing a margin for the probability of an erroneous guess, this scheme 
interpolates the two standard discrimination schemes: minimum-error and unambiguous 
discrimination. We present solutions for two cases. One is the case of two 
unitary processes with general prior probabilities. The other is the case with a 
group symmetry: The processes comprise a projective representation of a 
finite group. In the latter case, we found that unambiguous discrimination is a kind of 
``all or nothing'': The maximum success probability is either 0 or 1. 
We also thoroughly analyze how entanglement with an auxiliary system improves discrimination 
performance. 
\end{abstract}

\maketitle

\newcommand{\ket}[1]{|\,{#1}\,\rangle}
\newcommand{\bra}[1]{\langle\,{#1}\,|}
\newcommand{\braket}[2]{\langle\,#1\,|\,#2\,\rangle}
\newcommand{\mbold}[1]{\mbox{\boldmath $#1$}}
\newcommand{\sbold}[1]{\mbox{\boldmath ${\scriptstyle #1}$}}
\newcommand{\tr}{\,{\rm tr}\,}
\newcommand{\trm}{{\rm tr}}

\newtheorem*{theorem}{Theorem}

\section{Introduction} 
Suppose Alice performs some operation on a quantum system. 
How can Bob guess what operation Alice has done? 
Let us assume Bob can prepare the system in any initial state before Alice's 
operation. What Bob can do then is to perform some measurement on the system 
after Alice's operation, and guess her operation from the outcome of 
his measurement. This is the problem of discrimination of quantum processes, and 
it involves two fundamental issues in quantum information theory. 

First of all, quantum measurement is statistical in nature, 
and generally it destroys the state to be measured. It is, therefore, a nontrivial 
problem for Bob to find the best measurement to discriminate between generally 
nonorthogonal states after the operation.  
This is the problem of quantum state discrimination \cite{Helstrom76,Holevo82,Chefles00}. 
The second issue is entanglement. Bob should prepare the system in the optimal initial state 
so that his measurement will be most effective. One question is whether and how much the  
performance of discrimination improves by an input state entangled with an 
auxiliary system (ancilla). Another problem concerns the usefulness of an entangled 
input state into the parallel arrangement of processes of the same kind. 

The problem of discrimination of processes has received much attention in recent years, 
and a number of results have been reported. 
It has been shown that two distinct unitary devices can be perfectly 
discriminated by a finite number of devices arranged in parallel and an 
appropriate entangled input state \cite{Acin01,DAriano01}.  
Discrimination between unitary processes with a group symmetry has also been studied. 
Owing to the symmetry of the set of processes, one can determine the optimal measurement 
scheme and discrimination success probability in terms of group representations 
\cite{Chiribella04,Chiribella_J04,Chiribella05,Korff04,Hayashi05}. 
Thorough analyses of the asymptotic behavior in Lie group estimation can be found 
in Refs.~\cite{Lie_group_estimation,Imai09}. 
Quite recently, a necessary and sufficient condition for when general quantum operations are 
perfectly discriminated within a finite number of queries was reported \cite{Duan09}. 

In this paper, we assume Alice's operations are unitary. With given prior probabilities, 
she selects one from a finite set of unitary operations. The set of unitary 
operations and the prior probabilities are known to Bob. 
For state discrimination, two schemes have been extensively studied: 
minimum-error discrimination \cite{Helstrom76} and unambiguous discrimination which allows an 
inconclusive result \cite{Ivanovic87,Dieks88,Peres88,Jaeger95}. 
Recently, a new scheme has been proposed \cite{Touzel07,Hayashi08,Sugimoto09}, which 
interpolates the two standard schemes by introducing  
a margin on the mean probability of an erroneous guess. We adopt this scheme for 
process discrimination. 

We will thoroughly analyze two solvable cases. 
The first is the case of two unitary processes with arbitrary prior probabilities. 
Then, we examine the set of processes with group symmetry: The processes 
comprise a unitary projective representation of a finite group, and prior probabilities 
are equal. The maximum discrimination success probability is given in terms of 
dimensions and multiplicities of irreducible representations. 
We will clarify whether and how much 
the discrimination performance can be improved by an input state 
entangled with an ancilla system. 
   
\section{Unitary-process discrimination with error margin}
Suppose $n$ unitary operations $\{U_i\}_{i=1}^{n}$ are defined on a quantum system and 
Alice performs one of the operations $U_i$ with a prior probability $\eta_i$. 
Bob does not know which operation is performed by Alice though  
he has the knowledge of the set of operations and the prior probabilities. 
Bob's task is to optimally guess which operation was performed by Alice.
Bob can prepare the quantum system in any state (input state for the process) 
before Alice's operation. He can also perform any measurement on the quantum system 
after Alice's operation. This is the problem of unitary-process discrimination. 

Let us impose an error margin on the mean probability of Bob's incorrect guess. 
This is possible by allowing Bob to declare an inconclusive result ``I don't know." 
The positive-operator valued measure (POVM) of Bob's measurement consists of $n+1$ 
elements, $E_\mu\ (\mu=0,\ldots,n)$, where measurement outcome $1 \le \mu \le n$ means 
the process is identified with $U_\mu$, and $E_{\mu=0}$ produces the inconclusive result.  
By $P_{U_i,E_\mu}$ we denote the joint probability 
that the process is $U_i\ (i=1,\ldots,n)$ and the measurement outcome is 
$\mu\ (\mu=0,1,\ldots,n)$. 
The probability $P_{U_i,E_\mu}$ is given by 
\begin{align}
  P_{U_i,E_\mu} = \eta_i \tr U_i \rho U_i^\dag E_\mu, 
\end{align}
where $\rho$ is the input state chosen by Bob, which is generally mixed. 
The success probability of discrimination is then given by 
\begin{align}
  P_\circ \equiv \sum_{i=1}^n P_{U_i,E_i}. 
          \label{eq:p_circ}
\end{align} 
The mean probability of error is 
\begin{align} 
  P_\times \equiv \sum_{i,j=1\ (i \ne j)}^n P_{U_i,E_j}. 
          \label{eq:p_times}
\end{align}
We impose a margin $m$ on the mean probability of error, 
\begin{align}
   P_\times \le m.
\end{align}
Bob's task is to maximize the success probability $P_\circ$ subject to the constraint 
$P_\times \le m$ by choosing the POVM $\{E_\mu\}_{\mu=0}^n$ and the input state $\rho$ 
in an optimal way. 

The input state can generally be mixed. As shown in the following, 
the optimal input state can be assumed to be a pure state for $m=1$ and 
$m=0$. For a general error margin, we can give a sufficient condition 
such that that the maximum success probability can be attained by a pure input 
state.  

Let us express a general mixed input state by 
\begin{align}
   \rho = \sum_a \lambda_a \ket{a}\bra{a},  \label{eq:rho} 
\end{align}
where $\{\lambda_a\}$ is a probability distribution. The discrimination success probability 
and the margin condition for the mean error probability are given by 
\begin{align}
 P_\circ &= \sum_a \lambda_a \left( 
              \sum_i \eta_i \bra{a} U_i^\dagger E_i U_i \ket{a} \right) \nonumber \\
         &\equiv \sum_a \lambda_a P(a), 
               \label{eq:mixed_p_circ} \\
 P_\times &= \sum_a \lambda_a \left(
              \sum_{i \ne j} \eta_i \bra{a} U_i^\dagger E_j U_i \ket{a} \right) \nonumber \\ 
          &\equiv \sum_a \lambda_a m(a) \le m.
               \label{eq:mixed_p_times}
\end{align}
Here, $P(a)$ and $m(a)$ are the success probability and the error probability, respectively, 
when the input state is given by $\ket{a}$. 
Among the pure states $\ket{a}$ in Eq.~(\ref{eq:rho}), let $\ket{a_{\max}}$ be 
the pure state that has the greatest $P(a)$. 
When $m=1$, we can take $\ket{a_{\max}}$ for the input state, since the error-margin condition 
is inactive in this case. 
When $m=0$, we can also take $\ket{a_{\max}}$ for the input state, since 
Eq.~(\ref{eq:mixed_p_times}) implies all states $\ket{a}$ 
satisfy the no-error condition, $m(a)=0$. Thus, the input state can be assumed pure when 
$m=1$ or $0$. 

For the general error margin, consider the discrimination problem in which the input state is 
restricted to be pure, and denote the maximum success probability  
by $P_{\max}^{\text{pure}}(m)$. 
It is evident that $P_{\max}^{\text{pure}}(m)$ is a monotonically increasing function of $m$. 
Note that the inequality $P(a) \le P_{\max}^{\text{pure}}(m(a))$ holds by definition. 
Now, assume that $P_{\max}^{\text{pure}}(m)$ is a concave function of $m$. Then we observe 
\begin{align*}
  P_\circ = \sum_a \lambda_a P(a) 
        &\le  \sum_a \lambda_a P_{\max}^{\text{pure}}(m(a))  \\
        &\le  P_{\max}^{\text{pure}}(\sum_a \lambda_a m(a)) \\
        &\le  P_{\max}^{\text{pure}}(m). 
\end{align*}   
This implies that the success probability of any mixed input state never exceeds 
the success probability of the optimal pure input state. Thus, the concavity of 
$P_{\max}^{\text{pure}}(m)$ is a sufficient condition for the maximum success 
probability to be able to be attained by a pure input state. 

This argument holds regardless of the use of an auxiliary system.  
If we allow a sufficiently large auxiliary system for the input state 
and measurement, we can show stronger results: 
$P_{\max}(m)=P_{\max}^{\text{pure}}(m)$ and  $P_{\max}(m)$ is concave.
Suppose the unitary operations $U_i$ act on system $Q$ and we have an auxiliary system $R$. 
Assume the maximum success probability is attained by a generally mixed input state $\rho^{QR}$ 
and a POVM $E^{QR}_\mu$, which are defined on the composite system $QR$. 
Introducing another auxiliary system $S$, we consider purification of  $\rho^{QR}$, 
which we denote by $\ket{\Psi^{QRS}}$. If we take $\ket{\Psi^{QRS}}$ as input and measure 
the output by POVM $E_{\mu}^{QRS} \equiv E_{\mu}^{QR} \otimes \mbold{1}^S$, the 
success and error probabilities do not change. Thus, the optimality can always be achieved 
by a pure-state input if a sufficiently large ancilla can be used. 

The concavity of $P_{\max}(m)$ can be shown in the following way: 
Take two values of error margin, $m_1$ and $m_2$. For each error margin $m_j\, (j=1,2)$, 
we assume the set of input state $\rho_j^{QR}$ and POVM $E_{\mu}^{QR}(j)$ is optimal.  
Here, we introduce a two-dimensional auxiliary system $S$ with an orthonormal basis 
$\ket{1^S}$ and $\ket{2^S}$. 
In the composite system $QRS$, let us consider the input state 
\begin{align*}
 \rho^{QRS}  =  \lambda_1 \rho_1^{QR} \otimes \ket{1^S}\bra{1^S} 
                     +\lambda_2 \rho_2^{QR} \otimes \ket{2^S}\bra{2^S} , 
\end{align*}
where $\lambda_1,\lambda_2 \ge 0$ and $\lambda_1+\lambda_2 =1$, and POVM defined by 
\begin{align*}
  E_{\mu} ^{QRS} =  E_{\mu}^{QR}(1) \otimes \ket{1^S}\bra{1^S} 
                         +E_{\mu}^{QR}(2) \otimes \ket{2^S}\bra{2^S}. 
\end{align*} 
It is clear that the mean error probability is given by 
$P_\times = \lambda_1 m_1 + \lambda_2 m_2$.  
The mean success probability is also given by a similar form, 
$P_\circ = \lambda_1 P_{\max}(m_1) + \lambda_2 P_{\max}(m_2) $, 
which should not exceed the maximum success probability 
$P_{\max}(\lambda_1 m_1 + \lambda_2 m_2)$. This establishes the concavity of $P_{\max}(m)$ 
when  a sufficiently large ancilla is available.

\section{Discrimination between two unitary processes} 
In this section, we consider discrimination with error margin between two unitary processes 
$U_1$ and $U_2$ with prior probabilities $\eta_1$ and $\eta_2$, respectively. 

First, we assume the input state is fixed to be a certain pure state $\ket{\phi}$.  
Optimization is performed only with respect to POVM. Then, the problem reduces to discrimination 
between two pure states $\ket{\phi_1} \equiv U_1 \ket{\phi}$ and 
$\ket{\phi_2} \equiv U_2 \ket{\phi}$. This problem has already been solved in 
\cite{Hayashi08,Sugimoto09}. One of the three types of measurement is optimal 
depending on the following parameters: prior probabilities $\eta_i$, the inner product between the 
two states, and error margin $m$. 
The parameter space is divided into the following three domains: 
\begin{equation*}
    \begin{cases} 
      \text{minimum-error domain : } & m_c \le m \le 1, \\
      \text{intermediate domain : }  & m_c' \le m \le m_c,  \\
      \text{single-state domain : }  & 0 \le m \le m_c', 
    \end{cases}
\end{equation*}
where two critical error margins $m_c$ and $m_c'$ are defined by 
\begin{align}
    m_c & \equiv  \frac{1}{2}\left( 1 - \sqrt{1-4\eta_1\eta_2 S} \right),  
                   \label{eq:mc} \\
    m_c'& \equiv  
            \begin{cases} 
               \displaystyle
               \frac{(\eta_1-\sqrt{\eta_1\eta_2S})^2}{1-2\sqrt{\eta_1\eta_2S}} &
                            (\eta_1 \le \eta_2 S), \\
               0 & (\eta_1 \ge \eta_2 S).
            \end{cases}
                   \label{eq:mc_prime}
\end{align}
where $\eta_1 \le \eta_2$ is assumed and $S \equiv |\braket{\phi_1}{\phi_2}|^2$.  
In the minimum-error domain, the optimal measurement is the same as the one 
of minimum-error discrimination, which does not produce the inconclusive 
result, ``I don't know.'' In the single-state domain, one of the two states 
is omitted in the optimal measurement. In the intermediate domain, the probabilities 
for three measurement outcomes (the states $\ket{\phi_1}$ and $\ket{\phi_2}$ and the inconclusive 
result) are nonzero. 
The maximum success probability as a function of $m$ and $S$ is given by
\begin{align}
  & P_{\max}^{\text{pure}}(m,S) =  \nonumber \\  
  &  \begin{cases} 
        \frac{1}{2}\left( 1 + \sqrt{1-4\eta_1\eta_2 S} \right) &
                            (m_c \le m \le 1), \\ 
        \left( \sqrt{m} + \sqrt{ 1 - 2\sqrt{\eta_1\eta_2 S} } \right)^2  &
                            (m_c' \le m \le m_c), \\ 
        \eta_2 \left( \sqrt{\frac{m}{\eta_1}S} + \sqrt{\frac{\eta_1-m}{\eta_1}(1-S)} \right)^2 &  
                            (0 \le m \le m_c'). 
     \end{cases}
           \label{eq:maximum_p_circ}
\end{align}
See Refs.~\cite{Hayashi08,Sugimoto09} for details. 
It can be readily shown that $P_{\max}^{\text{pure}}(m,S)$ is a concave and monotonically 
increasing function of $m$. It is also evident that $P_{\max}^{\text{pure}}(m,S)$ 
is monotonically decreasing as a function of $S$. 

We can now optimize the success probability with respect to the input pure state $\ket{\phi}$ 
in the following way: 
\begin{align}
  P_{\max}^{\text{pure}}(m) &= \max_{\ket{\phi}} 
            P_{\max}^{\text{pure}}(m,|\bra{\phi}U_1^\dagger U_2\ket{\phi}|^2) \nonumber \\ 
                            &= P_{\max}^{\text{pure}}(m,S_{\min}), 
\end{align}
where $S_{\min}$ is defined to be 
\begin{align}
  S_{\min} \equiv \min_{\ket{\phi}} \left| \bra{\phi}U_1^\dagger U_2\ket{\phi} \right|^2.  
\end{align}
Since $P_{\max}^{\text{pure}}(m)=P_{\max}^{\text{pure}}(m,S_{\min})$ is concave for $m$, 
we conclude that the maximum success probability of discriminating two unitary processes 
can be attained by a pure state input; namely, $P_{\max}(m)=P_{\max}^{\text{pure}}(m)$. 

\begin{figure}
\includegraphics[width=0.8\hsize]{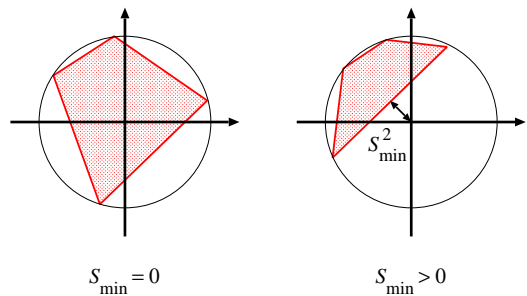}
\caption{\label{fig:smin}
(Color online) $S_{\min}$ and the convex polygon on the complex plane which is the convex 
hull of the eigenvalues of $U_1^{\dagger}U_2$. 
}
\end{figure}

$S_{\min}$ can be determined by eigenvalues of $U_1^{\dagger}U_2$ \cite{Acin01,DAriano01}. 
Let $\{e^{i\theta_1}, e^{i\theta_2}, \ldots, e^{i\theta_d} \}$ be eigenvalues of 
$U_1^{\dagger}U_2$, where $d$ is the dimension of the space considered. 
We can express $S_{\min}$ as 
\begin{align}
  S_{\min} &= \min_{\ket{\phi}} \left| \sum_{a=1}^d |\braket{\phi}{a}|^2 e^{i\theta_a} \right|^2 
                \nonumber \\ 
      &= \min_{q_a \ge 0, \sum_a q_a = 1} \left| \sum_{a=1}^d q_a e^{i\theta_a} \right|^2,  
\end{align}
where $\ket{a}$ is the eigenstate of $U_1^{\dagger}U_2$ with eigenvalue $e^{i\theta_a}$, 
and $q_a$ is defined to be $|\braket{\phi}{a}|^2$. 
We note that $\sum_{a=1}^d q_a e^{i\theta_a}$ with $q_a \ge 0, \sum_a q_a = 1$ 
is the convex hull of the points 
$\{e^{i\theta_1}, e^{i\theta_2}, \ldots, e^{i\theta_d} \}$ on the complex plane, and 
represents a convex polygon on the plane. If this polygon contains the origin, then 
$S_{\min} =0$, and consequently we obtain $P_{\max}(m)=1$. Otherwise, $S_{\min}$ is given by  
the square of the minimum distance between the polygon and the origin (see Fig. \ref{fig:smin}). 

Before concluding the section, we examine whether some use of an auxiliary 
system (ancilla) can help Bob improve the discrimination success probability. 
Suppose Alice's two unitary operations act on the system $Q$ alone, and Bob can prepare a 
certain input state in the composite system $QR$ and perform any measurement 
on $QR$ after Alice's operation, where $R$ is an auxiliary system. 
The set of eigenvalues of the operator $U_1^\dagger U_2$ is unchanged 
with only their multiplicities increased by a factor of the dimension of system $R$.   
This does not change $S_{\min}$. Thus, entanglement with an auxiliary system 
does not help in discrimination between two unitary processes. 
This contrasts with the case of discrimination discussed in the next section, 
which involves more than two unitary processes.   
 
\section{Unitary processes as a projective representation of a finite group}
It is generally hard to analyze a process discrimination problem of more than two processes. 
However, if the set of processes has some symmetry, the problem can be tractable. 
In this section, we consider a set of processes with a group symmetry. 
There are many interesting cases where a set of operations itself forms a group. 
Here, we consider a slightly generalized situation: The set of unitary processes 
$\{U_g\}_{g \in G}$ is assumed to be a unitary projective representation of a finite group $G$.  

More precisely, the set of unitary processes $\{U_g\}_{g \in G}$ satisfies
\begin{align}
      U_gU_h = c_{g,h} U_{gh}\ \ (g,h \in G),  \label{eq:projective_representation}
\end{align}
where $c_{g,h}$ is a complex number with $|c_{g,h}|=1$. 
The occurrence probabilities of process $U_g$ are assumed to be equal: 
$\eta_g = \frac{1}{|G|}$ with $|G|$ being the order of $G$.  

When all $c_{g,h}$ are 1, the projective representation reduces to an ordinary 
unitary representation of $G$. The set of complex numbers $\{c_{g,h}\}$ is 
called a factor set and satisfies the following cocycle conditions: 
\begin{align}
   c_{g,h}c_{gh,k} = c_{g,hk}c_{h,k}\ \ (g,h,k \in G), 
\end{align}
which is a consequence of the associativity of multiplication, $U_gU_hU_k$. 
The factor set depends on phase factors of each $U_g$. It is important that the factor 
set does not always reduce to a trivial one (all $c_{g,h}$ are 1) by redefining phase 
factors of $U_g$.  
A simple example is the set 
$\{\mbold{1}, \sigma_x, \sigma_y, \sigma_z\}$, where $\sigma_x$, $\sigma_y$, and 
$\sigma_z$ are Pauli matrices.
This set is not a group, but a projective representation of group $\mathbb{Z}_2 \times \mathbb{Z}_2$. 
We will later discuss a generalization of this example in connection with  superdense coding.  

For a general theory of projective representation of group (also known as ray representation) 
see, for example,  Ref.~\cite{Inui90}.
It is known that, for a fixed factor set, equivalence, 
reducibility, and irreducibility can be defined in the same way as in ordinary representations. 
Schur's lemma and the orthogonality relations of irreducible representation matrices also hold 
for projective representations.  

Further, in most of our calculations involving $U_g$, the factor set does not 
explicitly show up.  For example, the relation 
$U_g U_h A U_h^\dagger U_g^\dagger = U_{gh} A U_{gh}^\dagger$ 
still holds for any operator $A$. We also observe $U_1 A U_1^\dagger = A$ and 
$U_g A U_g^\dagger = U_{g^{-1}}^\dagger A U_{g^{-1}}$, where ``$1$'' represents the 
identity element of $G$. This is because, up to a phase factor, $U_1 = \mbold{1}$ and 
$U_g^\dagger = U_{g^{-1}}$.  

A POVM $\{E_0,E_g\}$ is said to be covariant \cite{Holevo82}, if it satisfies 
\begin{align*}
   U_g E_0 U_g^\dagger = E_0,\ U_h E_g U_h^\dagger = E_{hg}\ \ (g,h \in G). 
\end{align*} 
We can show that the optimal POVM can be assumed to be covariant. This useful property is not 
hampered by the error-margin condition and the factor set in the projective representation. 
Let $\{F_0,F_g\}$ be a POVM, which is not generally covariant. Construct another POVM 
$\{E_0,E_g\}$ as follows: 
\begin{align}
  E_0 = \frac{1}{|G|} \sum_{g \in G} U_g F_0 U_g^\dagger,\ \ E_g = U_g E_1 U_g^\dagger, 
\end{align}
where $E_1 \equiv \frac{1}{|G|} \sum_{g \in G} U_g^\dagger F_g U_g$.   
It is evident that $\{E_0,E_g\}$ is a covariant POVM. We find that the two POVMs 
give the same success probability. 
\begin{align*}
  P_\circ(F) &\equiv \frac{1}{|G|}\sum_{g \in G} \tr F_g U_g \rho U_g^\dagger 
                 = \tr E_1 \rho  \\
             &= \frac{1}{|G|}\sum_{g \in G} \tr E_g U_g \rho U_g^\dagger 
                 \equiv P_\circ(E).
\end{align*}
The error probability is also the same for the two POVMs. 
\begin{align*}
  P_\times(F) &\equiv \frac{1}{|G|}\sum_{g \ne h} \tr F_g U_h \rho U_h^\dagger 
                 = \sum_{g(\ne 1)}\tr E_1 U_g \rho U_g^\dagger  \\
             &= \frac{1}{|G|}\sum_{g \ne h} \tr E_g U_h \rho U_h^\dagger 
                 \equiv P_\times(E).
\end{align*}
Thus, if a POVM $\{F_0,F_g\}$ is optimal, so is the covariant POVM $\{E_0,E_g\}$ 
which is constructed from $\{F_0,F_g\}$. 

In terms of the covariant POVM, the task is to maximize 
\begin{align}
   P_\circ = \tr E_1 \rho, 
\end{align}
subject to conditions 
\begin{align}
   & E_1 \ge 0,\ \ \sum_{g \in G} U_g E_1 U_g^\dagger \le 1, \\
   & P_\times = \sum_{g \ne 1} \tr E_1 U_g \rho U_g^\dagger \le m, 
\end{align}
where variables are POVM element $E_1$ and input state $\rho$.    

\subsection{Case of irreducible representation}
Let us assume the representation $U_g$ is irreducible and determine 
the maximum success probability $P_{\max}(m)$. We will see that this 
simple case provides a helpful guide line for the more general case considered in the 
next section. 

Define an operator $A$ to be $\sum_{g \in G} U_g E_1 U_g^\dagger$. It can be 
readily shown that $A$ commutes with $U_g$ for all $g$ in $G$. 
According to Schur's lemma, operator $A$ is the identity up to a 
factor, since the representation $U_g$ is irreducible. The factor can be fixed by 
calculating traces. We find 
\begin{align}
   \sum_{g \in G} U_g E_1 U_g^\dagger = \frac{|G|\tr E_1}{d} \mbold{1}, 
         \label{eq:A}
\end{align} 
where $d$ is the dimension of the space considered. The POVM element $E_1$ is positive 
semidefinite, and it clearly satisfies $\tr(E_1)\mbold{1} \ge E_1$. Combining this 
inequality and Eq.~(\ref{eq:A}), we obtain 
\begin{align}
   E_1 \le \frac{d}{|G|} \sum_{g \in G} U_g E_1 U_g^\dagger , 
       \label{eq:key_inequality_irreducible}
\end{align}
which serves as a key inequality for determining $P_{\max}(m)$. 

Using this inequality, we derive two upper bounds for the success probability 
$P_\circ$. The first upper bound is obtained in the following way:
\begin{align}
  P_\circ = \tr E_1 \rho \le \frac{d}{|G|}  
          \tr \left( \sum_{g \in G} U_g E_1 U_g^\dagger \rho \right)
          \le \frac{d}{|G|},   
                  \label{eq:first_inequality}
\end{align} 
where the condition 
\begin{align*}
 \sum_{g \in G} U_g E_1 U_g^\dagger \le \mbold{1}, 
\end{align*}
is used in the last inequality. This upper bound is 
independent of the error margin. To obtain another upper bound involving the error margin, we 
slightly rewrite the inequality of Eq.~(\ref{eq:key_inequality_irreducible}) as
\begin{align*}
  E_1 \le \frac{d}{|G|} \left( E_1 + \sum_{g \ne 1} U_g E_1 U_g^\dagger \right),
\end{align*}
which immediately leads to 
\begin{align}
  E_1 \le \frac{1}{\frac{|G|}{d}-1} \sum_{g \ne 1} U_g E_1 U_g^\dagger.
          \label{eq:key_inequality_irreducible_prime}
\end{align}
We note that $|G| \ge d$ for an irreducible representation and equality occurs only for 
a trivial case $|G|=1$. Hereafter, we assume $|G| > 1$. 
Using this inequality, we obtain the second upper bound as 
\begin{align}
    P_\circ &= \tr E_1 \rho \le \frac{1}{\frac{|G|}{d}-1} 
                \tr \left( \sum_{g \ne 1} U_g E_1 U_g^\dagger \rho \right)   
                        \nonumber \\
          & = \frac{1}{\frac{|G|}{d}-1} P_\times  
           \le \frac{m}{\frac{|G|}{d}-1}.  
                  \label{eq:second_inequality}
\end{align} 

Combining the two upper bounds, we have 
\begin{align}
  P_\circ & \le 
     \begin{cases}
          \frac{d}{|G|} & \left( 1-\frac{d}{|G|} \le m \le 1 \right), \\
          \frac{m}{\frac{|G|}{d}-1} & \left( 0 \le m <  1-\frac{d}{|G|} \right), 
     \end{cases}
             \nonumber \\
          & \equiv f(m). 
\end{align}  

It is readily verified that this upper bound can be attained by any pure state input 
$\rho=\ket{\phi}\bra{\phi}$ and the POVM element $E_1 = f(m) \ket{\phi}\bra{\phi}$.  
Thus, the maximum success probability in the case of an irreducible representation 
is given by
\begin{align}
   P_{\max}(m) = P_{\max}^{\text{pure}}(m) = 
     \begin{cases}
          \frac{d}{|G|} & \left( m_c \le m \le 1 \right), \\
          \frac{m}{\frac{|G|}{d}-1} & \left( 0 \le m < m_c \right),   
     \end{cases}
\end{align}
where the critical error margin $m_c$ is defined as
\begin{align}
   m_c = 1-\frac{d}{|G|}.
\end{align}

If $m \ge m_c$, $P_{\max}(m)$ is given by that of minimum-error discrimination. 
Below $m_c$, $P_{\max}(m)$ is linear in error margin $m$. 
Interestingly, we find that $P_{\max}(0)=0$ unless $|G|=d$, implying 
it is impossible to unambiguously discriminate processes $U_g$ unless $|G|=d$. 
If $|G|=d$ instead, we find $P_{\max}(0)=1$. Thus, $P_{\max}(0)$ is either 0 or 1. 
Rather surprisingly, we will see that these features of the irreducible case are 
preserved in the more general case discussed next.     

\subsection{General case}
Here, we consider the representation $U_g$ to be generally reducible. 
When $U_g$ is irreducible, the inequality (\ref{eq:key_inequality_irreducible}) 
was essential to determination of the maximum success probability. 
Though we cannot resort to Schur's lemma as in the preceding section,    
there exists a generalization of Eq.~(\ref{eq:key_inequality_irreducible})  
for generally reducible representations. In this respect, we can show that 
the following general theorem holds: 

\begin{theorem}
Let $\{U_g\}_{g \in G}$ be a unitary projective representation of a finite group $G$
of order $|G|$. 
Define constant $\kappa$ as 
\begin{align} 
   \kappa \equiv \sum_\sigma \frac{\min(m_\sigma,d_\sigma) d_\sigma}{|G|}, 
            \label{eq:kappa}
\end{align}   
where $\sigma$ represents each irreducible representation of $G$, and $d_\sigma$ and $m_\sigma$ 
are the dimension and the multiplicity of irreducible representation $\sigma$ in the decomposition 
of $U_g$, respectively.  
Then, for any positive semidefinite operator $E$, the following inequality holds:
\begin{align}
   E \le \kappa \sum_{g \in G} U_g E U_g^\dagger. \label{eq:key_inequality}
\end{align}
\end{theorem}

The quantity $d_\sigma^2/|G|$ is called the Plancherel measure of irreducible 
representation $\sigma$. It is known that they sum to unity when summed over all possible 
irreducible representations for a given factor set \cite{Inui90}. 
Thus, the constant $\kappa$ is generally less than or equal to 1.  
Note that if $U_g$ is irreducible, the generalized inequality Eq.~(\ref{eq:key_inequality}) 
reduces to Eq.~(\ref{eq:key_inequality_irreducible}), 
since $\kappa$ is then given by $d/|G|$.   

Before proving the theorem, we introduce a representation basis and 
representation matrices. Decomposing the representation $U_g$ 
into irreducible representations, we obtain an orthonormal basis written  
as $\ket{\sigma,b,a}\ (a=1,\ldots,d_\sigma,b=1,\ldots,m_\sigma)$. 
Here, $\sigma$ represents each irreducible representation of $G$. 
Index $a$ specifies each vector belonging to irreducible 
representation $\sigma$, and $a$ runs from $1$ to $d_\sigma$. Index $b$ 
stands for ``other quantum numbers,'' which are invariant under any operation $U_g$. 
Index $b$, therefore, runs from $1$ to the multiplicity $m_\sigma$. 
The basis states $\ket{\sigma,b,a}$ transform under operation $U_g$ as follows:
\begin{align}
 U_g \ket{\sigma,b,a}  
   &=\sum_{a'=1}^{d_\sigma} \ket{\sigma,b,a'}\bra{\sigma,b,a'}U_g\ket{\sigma,b,a}
                   \nonumber \\ 
   &=\sum_{a'=1}^{d_\sigma} D^{\sigma}_{a'a}(g) \ket{\sigma,b,a'}. 
      \label{eq:transformation} 
\end{align} 
Here, $D^{\sigma}_{a'a}(g) \equiv \bra{\sigma,b,a'}U_g\ket{\sigma,b,a}$ 
are irreducible representation matrices, and that are known to satisfy the 
following orthogonal relations: 
\begin{align}
  \sum_{g \in G} D^{\sigma *}_{a_1a_2}(g)D^{\sigma'}_{a_1'a_2'}(g)
  = \delta_{\sigma\sigma'}\delta_{a_1a_1'}\delta_{a_2a_2'} \frac{|G|}{d_\sigma}.
         \label{eq:orthogonality}
\end{align}

We now present the proof of the theorem. 
\begin{proof}
Any positive semidefinite operator $E$ can be written as 
\begin{align}
   E = \sum_e \ket{e}\bra{e}, 
\end{align}
where $\ket{e}$ is not generally normalized. If each term in this expression satisfies 
inequality (\ref{eq:key_inequality}), so does $E$. 
Thus, it suffices to prove the case in which the rank of $E$ is one: 
 $ E = \ket{e}\bra{e} $. 
The vector $\ket{e}$ can be expanded in terms of the basis $\ket{\sigma,b,a}$ as 
\begin{align}
  \ket{e} = \sum_\sigma \sum_{b=1}^{m_\sigma} \sum_{a=1}^{d_\sigma} 
             e_{\sigma b a} \ket{\sigma,b,a}. \nonumber
\end{align}
Here, it is convenient to redefine the basis $\ket{\sigma,b,a}$ so that 
the coefficient $e_{\sigma b a}$ is diagonal with respect to $b$ and $a$. 
This is possible by the singular value decomposition of the matrix with $(b,a)$ 
entry given by $e_{\sigma b a}$. Note that 
the transformation property of Eq.~(\ref{eq:transformation}) is unchanged by 
this redefinition. In this redefined basis, we can write 
\begin{align}
  \ket{e} = \sum_\sigma \sum_{a=1}^{\tilde d_\sigma} 
             e_{\sigma a} \ket{\sigma,a,a},  
\end{align}
where $\tilde d_\sigma \equiv \min(m_\sigma,d_\sigma)$. 
Using the orthogonality of irreducible matrices given in Eq.~(\ref{eq:orthogonality}), 
we obtain 
\begin{align}
  & \sum_{g \in G} U_g E U_g^\dagger =  \sum_{g \in G} U_g \ket{e}\bra{e} U_g^\dagger 
                  \nonumber \\
  & = \sum_\sigma \sum_{a=1}^{\tilde d_\sigma} \frac{|G|}{d_\sigma}|e_{\sigma a}|^2 
      \sum_{a'=1}^{d_\sigma} \ket{\sigma,a,a'}\bra{\sigma,a,a'}. 
    \label{eq:UEU}
\end{align}

Now, let $\ket{\phi}$ be an arbitrary state. 
Writing $\phi_{\sigma b a} \equiv \braket{\sigma,b,a}{\phi}$, we find  
\begin{align*}
 &  \bra{\phi}E\ket{\phi} = |\braket{\phi}{e}|^2 
   = \left| \sum_\sigma \sum_{a=1}^{\tilde d_\sigma} e_{\sigma a} \phi_{\sigma a a}^* \right|^2  \\
 &= \left| \sum_\sigma \sum_{a=1}^{\tilde d_\sigma} 
                    \sqrt{\frac{d_\sigma}{|G|}} 
                    \sqrt{\frac{|G|}{d_\sigma}}e_{\sigma a}\phi_{\sigma a a}^*  \right|^2 \\
 &\le \left( \sum_\sigma \sum_{a=1}^{\tilde d_\sigma} \frac{d_\sigma}{|G|} \right) 
         \left( \sum_\sigma \sum_{a=1}^{\tilde d_\sigma} \frac{|G|}{d_\sigma}|e_{\sigma a}|^2 
                                             |\phi_{\sigma a a}|^2 \right),  
\end{align*}
where the Schwarz inequality is used. The first factor in the last line is 
the constant $\kappa$ defined by Eq.~(\ref{eq:kappa}). 
To evaluate the second factor, we use Eq.~(\ref{eq:UEU}) and obtain 
\begin{align*}
   \bra{\phi} \sum_{g \in G} U_g E U_g^\dagger \ket{\phi}
   = \sum_\sigma \sum_{a=1}^{\tilde d_\sigma} \frac{|G|}{d_\sigma}|e_{\sigma a}|^2 
                 \sum_{a'=1}^{d_\sigma} |\phi_{\sigma a a'}|^2 ,
\end{align*}                                             
which is clearly greater than or equal to the second factor. 
Thus, we obtain
\begin{align*}
  \bra{\phi}E\ket{\phi} \le \kappa \bra{\phi} \sum_{g \in G} U_g E U_g^\dagger \ket{\phi}. 
\end{align*} 
Equality holds if and only if $e_{\sigma a}^* \phi_{\sigma a a}/d_\sigma$ is 
independent of $\sigma$ and $a$, and $\phi_{\sigma b a}=0$ for $b \ne a$. 
Since $\ket{\phi}$ is arbitrary, we obtain the desired result. 
This completes the proof of the theorem. 
\end{proof}

With the key inequality of Eq.~(\ref{eq:key_inequality}) at hand, we can determine 
the maximum success probability along the same lines as the irreducible case. 
The key inequality immediately leads to the first upper bound for the success probability  
\begin{align}
  P_\circ(m) \le \kappa. 
\end{align}
By rewriting the key inequality as in the irreducible case, we have  
\begin{align}
  E_1 \le \frac{\kappa}{1-\kappa} \sum_{g \ne 1} U_g E_1 U_g^\dagger, 
\end{align}
which is a generalization of Eq.~(\ref{eq:key_inequality_irreducible_prime}). 
The second upper bound follows from this inequality. 
\begin{align}
  P_\circ \le \frac{\kappa}{1-\kappa} m.
\end{align} 

Combining the two upper bounds, we have 
\begin{align}
  P_\circ &\le \begin{cases}
                       \kappa & (1-\kappa \le m \le 1) \\
                       \frac{\kappa}{1-\kappa}m & (0 \le m < 1-\kappa), 
                  \end{cases}  \\
              &\equiv f(m). \nonumber 
\end{align} 
This upper bound is attained by the following pure-state input: 
\begin{align}
   \ket{\phi} = \frac{1}{\sqrt{\kappa}} \sum_\sigma \sum_{a=1}^{\tilde d_\sigma} 
                   \sqrt{\frac{d_\sigma}{|G|}} \ket{\sigma,a,a}, 
\end{align}
and the POVM element $E_1$ of rank 1 given by 
\begin{align}
   E_1 = f(m) \ket{\phi}\bra{\phi}.
\end{align}

\begin{figure}
\includegraphics[width=0.7\hsize]{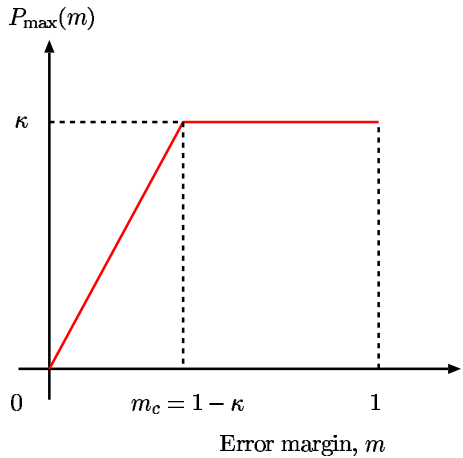}
\caption{\label{fig:representation} 
(Color online)  The maximum discrimination success probability $P_{\max}(m)$ of processes 
$\{U_g\}_{g \in G}$ which comprise a projective representation of a finite group $G$. 
$m$ denotes the margin for the mean error probability. The constant $\kappa$ is given 
in Eq.~(\ref{eq:kappa}), 
which is replaced by $\kappa^{\rm A}$ in Eq.~(\ref{eq:kappa_A}) when a sufficiently 
large ancilla system can be employed. 
}
\end{figure}

Thus, the maximum success probability $P_{\max}(m)$ is given by 
\begin{align}
     P_{\max}(m) = P_{\max}^{\text{pure}}(m)
            &= \begin{cases}
                       \kappa & (m_c \le m \le 1), \\
                       \frac{\kappa}{1-\kappa}m & (0 \le m < m_c), 
                  \end{cases} 
             \label{eq:pmax_representation}
\end{align} 
where $m_c = 1-\kappa$ and 
$\kappa = \sum_\sigma \frac{\min(m_\sigma,d_\sigma) d_\sigma}{|G|}$. 
For minimum-error discrimination ($m=1$), this maximum success probability reproduces 
the result obtained in \cite{Korff04}. 
As in the irreducible case, we find again that $P_{\max}(m)$ is linear below the 
critical error margin $m_c$ and reaches the probability of minimum-error discrimination 
at $m=m_c$ (see Fig.~\ref{fig:representation}). Unambiguous discrimination is again 
``all or nothing'': $P_{\max}(0)$ is either 0 or 1.   These features contrast with the case of 
discrimination between two unitary processes with no group symmetry.  Note that 
a set of two unitaries cannot always be considered as a projective representation of 
some group. 
This is because the group should be $\mathbb{Z}_2$ in this case,  and this would mean, 
up to a phase factor, the square of $U_1^{\dagger}U_2$  should be the identity.

\section{Entanglement with ancilla and illustrative examples}  
We examine how entanglement with an ancilla system improves the 
discrimination performance. Let $Q$ be the system on which the process $U_g^Q$ 
acts, and let $R$ be an ancilla system. We assume that the input state can be 
any (generally entangled) state $\ket{\phi}^{QR}$ of the composite system $QR$, and any 
measurement on system $QR$ can be performed after the operation of $U_g^Q$.   

Clearly, $U_g^Q \otimes \mbold{1}^R$ is a projective representation of $G$, and 
the arguments given in the preceding section can be applied to the composite system $QR$ 
as well. We write the basis of irreducible representations in $QR$ as 
\begin{align}
   \ket{\sigma,(b,r),a}^{QR} \equiv \ket{\sigma,b,a}^Q \otimes \ket{r}^R, 
\end{align}
where $\{\ket{r}\}$ is an orthonormal basis of $R$. By introducing the ancilla, 
the multiplicity of irreducible representation $\sigma$ is increased by a factor 
of the dimension of the space of $R$, which we denote by $|R|$. 
The maximum success probability is still given by Eq.~(\ref{eq:pmax_representation}),  
but with the constant $\kappa$ replaced by 
\begin{align}
  \kappa' = \sum_\sigma \frac{\min(m_\sigma|R|,d_\sigma) d_\sigma}{|G|}, 
\end{align}
and, for an ancilla of a sufficiently large dimension, by  
\begin{align}
  \kappa^{\rm A} = \sum_{\sigma(m_\sigma \ge 1)} \frac{d_\sigma^2}{|G|},  \label{eq:kappa_A}
\end{align}
which is the Plancherel measure of irreducible representations appearing in the 
representation $U_g^Q$ \cite{Hayashi05}. Note that $\kappa^{\rm A} =1$ if all irreducible 
representations are in the decomposition of $U_g$. 
Since $\kappa \le \kappa' \le \kappa^{\rm A}$, the success probability 
generally improves by an ancilla.  It should be noted that any ancilla 
is useless ($\kappa = \kappa^{\rm A}$) if all irreducible representations are 
one dimensional.  This applies to an ordinary representation of an Abelian group. 
For a nontrivial projective representation of an Abelian group, however,  this is not 
always true, which will be illustrated by an example later. 

Let us focus on the unambiguous discrimination case ($m=0$). Without an ancilla, 
the success probability is 1 when $m_\sigma \ge d_\sigma$ for all possible irreducible representations $\sigma$. Otherwise, it is 0. If an irreducible representation 
$\sigma$ is missing in the decomposition of $U_g^Q$, meaning $m_\sigma=0$, any ancilla 
does not help. This is because the missing representation does not appear with any ancilla.
The most interesting case is probably when no irreducible representation is missing 
($m_\sigma \ge 1$ for all $\sigma$), but $m_\sigma < d_\sigma$ for some $\sigma$. 
Then, the success probability without an ancilla is 0. 
With a sufficiently large ancilla, however, the maximum success probability becomes 1. 

In what follows, we present three examples, which illustrate some differences in 
usefulness of an ancilla system. 
\subsection{Phase shift discrimination} 
Consider the following phase shift processes on a qubit: 
\begin{align}
  \begin{cases} 
    U_k \ket{0} = \ket{0}, \\ 
    U_k \ket{1} = e^{i\frac{2\pi}{K}k} \ket{1}, \\
  \end{cases} 
  \ k=0,1,\ldots, K-1, 
\end{align}
where $K$ is a positive integer. $\{U_k\}_{k=0}^{K-1}$ is an ordinary representation of 
the Abelian group $\mathbb{Z}_K$. 
All irreducible representations of $\mathbb{Z}_K$ are one dimensional and specified by an integer 
$\sigma$ ($= 0,1,\ldots,K-1$) as $D^\sigma(k)=e^{i\frac{2\pi}{K}\sigma k}$. 
The representation $U_k$ contains two irreducible representations, $\sigma =0$ and $\sigma=1$. 
As mentioned, an ancilla is useless for any error margin in this example. We find that the maximum 
success probability is given by Eq.~(\ref{eq:pmax_representation}) with $\kappa=\kappa^A=2/K$. 

This example provides one of the cases in which we can easily calculate 
the maximum success probability   
when the operations are performed on $N$ identical systems in parallel, and the input state of the 
$N$ systems is allowed to be entangled among its subsystems. 
The processes are now expressed as $U_k^{\otimes N}$ for an $N$-qubit system. 
We observe 
\begin{align*}
    U_k^{\otimes N} \ket{b_1 b_2 \cdots b_N} = e^{i\frac{2\pi}{K}k(b_1+b_2+ \cdots + b_N)} 
                     \ket{b_1 b_2 \cdots b_N},  
\end{align*}   
which shows that each computational basis state of the $N$-qubit system is an irreducible 
representation, with $\sigma$ given by the number of entries of 1.  
Thus, for $N \le K-2$, we find  $\kappa=(N+1)/K$, and for $N \ge K-1$, we find $\kappa=1$.    

\subsection{Quantum color coding}
The second example is quantum color coding \cite{Korff04,Hayashi05}. 
Consider $N$ identical quantum systems, each defined on vector space $\mathbb{C}^d$. 
Suppose Alice randomly permutes the $N$ quantum systems. 
Bob's task is to identify which permutation was performed by Alice. 
The dimension $d$ can be interpreted as the number of colors, and $N$ as the number of 
colored boxes to be identified.  
Alice's operations comprise a set of $N!$ permutation processes $U_g$ on 
$(\mathbb{C}^d)^{\otimes N}$, which is an ordinary representation of the symmetric group of degree $N$. 

\begin{figure}
\includegraphics[width=\hsize]{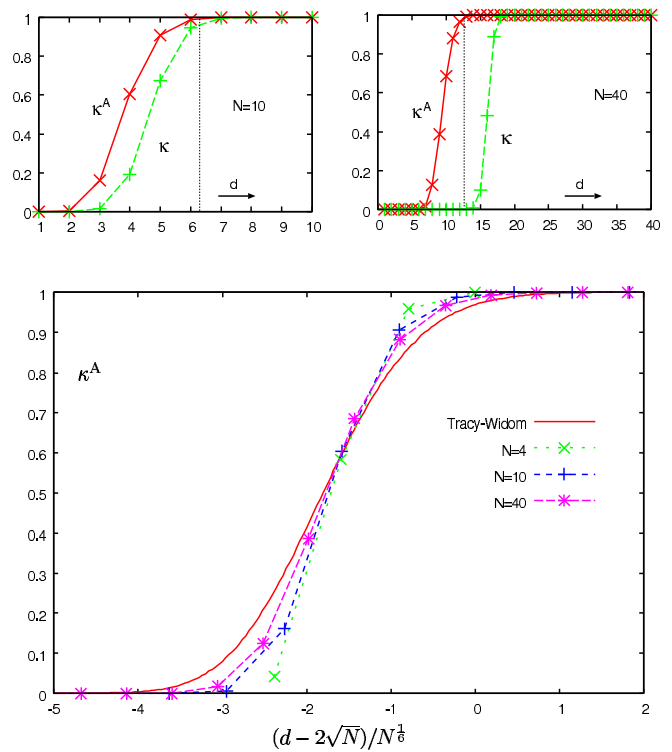} 
\caption{\label{fig_kappa} 
(Color online)  The constants $\kappa^{\rm A}$ and $\kappa$ in the quantum color coding, 
which are the maximum probabilities in minimum-error discrimination 
with and without an ancilla system, respectively. 
$N$ is the number of colored boxes, and $d$ is the number of colors. 
In the upper two figures, $\kappa$ and $\kappa^{\rm A}$ are compared. 
The vertical dotted lines indicate the positions of $d = 2\sqrt{N}$. 
In the bottom figure, $\kappa^{\rm A}$ are plotted 
as functions of $(d-2\sqrt{N})/N^{1/6}$, so that they approach the Tracy-Widom 
distribution in the large-$N$ limit \cite{Hayashi05}.  
}
\end{figure}

If $d \ge N$, it is clear that Bob can discriminate Alice's permutation with certainty. 
If $d < N$, $\kappa$ and $\kappa^{\rm A}$ are less than 1. For small $N$, differences  
between $\kappa$ and $\kappa^{\rm A}$ are not remarkable. 
For example, when $N=4$ and $d=2$, we find $\kappa=1/2$ and $\kappa^{\rm A}=7/12$. 
However, in the large-$N$ limit, $\kappa$ goes to 1 if $d > N/e$ \cite{Korff04}, 
and $\kappa^{\rm A}$ goes to 1 if $d > 2\sqrt{N}$ \cite{Hayashi05}. 
Thus, for $m>0$, entanglement with an ancilla improves the discrimination performance 
substantially (see Fig.~\ref{fig_kappa}). 

For unambiguous discrimination ($m=0$), however, ancilla does not help.  
If $d < N$, some irreducible representations, for example, the totally antisymmetric 
representation, are missing in the decomposition of $U_g$. Then, with 
any ancilla, Bob cannot unambiguously discriminate Alice's permutation. 

\subsection{Superdense coding}
The last example is the celebrated superdense coding in general dimensions \cite{Bennett92}. 
Consider the following unitary operators on $\mathbb{C}^d$: 
\begin{align}
    U_{k,l} = X^kZ^l\ \ (k,l= 0,\ldots,d-1),  \label{eq:proj_rep}
\end{align}
where $X$ and $Z$ are generalizations of Pauli matrices $\sigma_x$ and $\sigma_z$, 
respectively, given by 
\begin{align*}
   & X = \sum_{a=0}^{d-1} \ket{a}\bra{a+1}, \\
   & Z = \sum_{a=0}^{d-1} e^{i\frac{2\pi}{d}a}\ket{a}\bra{a}.
\end{align*} 
By the relation $XZ=e^{i\frac{2\pi}{d}}ZX$, we have 
\begin{align*}
   U_{k,l}U_{k',l'} = e^{-i\frac{2\pi}{d}lk'} U_{k+k',l+l'},
\end{align*}
which shows that $U_{k,l}$ is a projective representation of $\mathbb{Z}_d \times \mathbb{Z}_d$.  
The factor set is given by $c_{(k,l),(k',l')}= e^{-i\frac{2\pi}{d}lk'}$. 
Remember that equivalence of projective representations is defined for a fixed factor set. 
For this factor set, $U_{k,l}$ turns out to be the unique irreducible representation of 
$\mathbb{Z}_d \times \mathbb{Z}_d$. 
The dimension of this unique irreducible representation is $d$ and 
its multiplicity is 1, which gives $\kappa=\frac{1}{d}$ and $\kappa^{\rm A}=1$. 
Thus, without ancilla, the unambiguous discrimination probability is 0. 
However, we can perfectly discriminate the $d^2$ processes $U_{k,l}$ by using an ancilla 
system of dimension $d$. In fact, as is well known, the states 
$\ket{\phi_{k,l}}^{QR} = U_{k,l}^Q\otimes \mbold{1}^R \ket{\phi}^{QR}$ are mutually 
orthogonal if we take the following entangled input state:
\begin{align*}
  \ket{\phi}^{QR} = \frac{1}{\sqrt{d}} \sum_{a=0}^{d-1} \ket{a}^Q \otimes \ket{a}^R.  
\end{align*} 

This example clearly shows the difference between ordinary representations and nontrivial 
projective representations of the same group. Consider the following phase shift 
operations for a qutrit system: 
\begin{align}
  V_{k,l} = {\rm diag} \left( 1, e^{i\frac{2\pi}{d}k}, e^{i\frac{2\pi}{d}l} \right) 
             \ \ (k,l= 0,\ldots,d-1), 
                \label{eq:usual_rep}
\end{align}
which is an ordinary representation of the Abelian group $\mathbb{Z}_d \times \mathbb{Z}_d$.  
Any ancilla is useless for $V_{k,l}$, though $V_{k,l}$ and $U_{k,l}$ 
are both representations of the same group $\mathbb{Z}_d \times \mathbb{Z}_d$.

\section{Concluding Remarks} 
We have studied unitary-process discrimination with an error margin. 
By imposing a margin on the mean error probability, this scheme interpolates 
minimum-error and unambiguous discrimination. 

Two cases have been thoroughly analyzed and solutions were presented. 
One is the case of two unitary processes with arbitrary prior probabilities. 
The other is the set of processes with group symmetry: The processes 
comprise a unitary projective representation of a finite group, and prior probabilities 
are equal. Especially, in the latter case, we clarified the conditions 
under which discrimination performance improves by an input state 
entangled with an ancilla system. 
This analysis is quite general and applicable to many interesting unitary-process 
discrimination problems with a group symmetry. 
It will also be of interest in the future studies to extend our scheme to discrimination problems 
of isometries, some classes of a group, and quantum channels with group symmetry.

\begin{acknowledgments}
MH was partially supported by a Grant-in-Aid for Scientific Research in
the Priority Area ``Deepening and Expansion of Statistical Mechanical
Informatics'' (DEX-SMI),'No.~18079014, and a MEXT Grant-in-Aid for Young
Scientists (A), No.~20686026.
The Centre for Quantum Technologies is funded by the Singapore Ministry
of Education and the National Research Foundation as part of the
Research Centres of Excellence programme.
\end{acknowledgments}


\begin{thebibliography}{99}
\bibitem{Helstrom76}
C.~W.~Helstrom, 
{\it Quantum Detection and Estimation Theory} 
(Academic Press, New York, 1976). 

\bibitem{Holevo82} 
A.~S.~Holevo, 
{\it Probabilistic and statistical aspects of quantum theory} 
(North-Holland, Amsterdam, 1982).

\bibitem{Chefles00}
A.~Chefles, 
Contemp. Phys. {\bf 41}, 401 (2000). 

\bibitem{Acin01}
A.~Acin,
Phys. Rev. Lett. {\bf 87}, 177901 (2001). 

\bibitem{DAriano01}
G.~M.~D'Ariano, P.~Lo Presti, and M.~G.~A.~Paris, 
Phys. Rev. Lett. {\bf 87}, 270404 (2001). 

\bibitem{Chiribella04} 
G.~Chiribella, G.~M.~D'Ariano, P.~Perinotti, and M.~F.~Sacchi, 
Phys. Rev. A {\bf 70,} 062105 (2004). 

\bibitem{Chiribella_J04} 
G.~Chiribella and G.~M.~D'Ariano, 
J. Math. Phys. {\bf 45}, 4435 (2004). 

\bibitem{Chiribella05}  
G.~Chiribella, G.~M.~D'Ariano, and M.~F.~Sacchi,
Phys. Rev. A {\bf 72}, 042338 (2005).  

\bibitem{Korff04}
J.~Von Korff and J.~Kempe, 
Phys. Rev. Lett. {\bf 93}, 260502 (2004).

\bibitem{Hayashi05}
A.~Hayashi, T.~Hashimoto, and M.~Horibe, 
Phys. Rev. A {\bf 71}, 012326 (2005). 


\bibitem{Lie_group_estimation} 
A.~Peres and P.~F.~Scudo,
Phys. Rev. Lett. {\bf 87}, 167901 (2001);
E.~Bagan, M.~Baig, and R.~Munoz-Tapia, 
Phys. Rev. Lett. {\bf 87}, 257903 (2001); 
A.~Peres and P.~F.~Scudo, 
J. Mod. Opt. {\bf 49}, 1235 (2002); 
G.~Chiribella, G.~M.~D'Ariano, P.~Perinotti, and M.~F.~Sacchi, 
Phys. Rev. Lett. {\bf 93}, 180503 (2004); 
E.~Bagan, M.~Baig, and R.~Munoz-Tapia, 
Phys. Rev. A {\bf 70}, 030301(R) (2004);
M.~Hayashi,
Phys. Lett., A {\bf 354}, 183 (2006).

\bibitem{Imai09}
H.~Imai and M.~Hayashi, 
New J. Phys. {\bf 11}, 043034 (2009).


\bibitem{Duan09}
R.~Duan, Y.~Feng, and M.~Ying, 
Phys. Rev. Lett. {\bf 103}, 210501 (2009).

\bibitem{Ivanovic87}
I.~D.~Ivanovic,
Phys. Lett. A {\bf 123}, 257 (1987).

\bibitem{Dieks88}
D.~Dieks,
Phys. Lett. A {\bf 126}, 303 (1988).

\bibitem{Peres88} 
A.~Peres,
Phys. Lett. A {\bf 128}, 19 (1988). 

\bibitem{Jaeger95}
G.~Jaeger and A.~Shimony, 
Phys. Lett. A {\bf 197}, 83 (1995). 

\bibitem{Touzel07}
M.~A.~P.~Touzel, R.~B.~A.~Adamson, and A.~M.~Steinberg, 
Phys. Rev. A {\bf76}, 062314 (2007). 

\bibitem{Hayashi08}
A.~Hayashi, T.~Hashimoto, and M.~Horibe, 
Phys. Rev. A {\bf 78}, 012333 (2008).  

\bibitem{Sugimoto09}
H.~Sugimoto, T.~Hashimoto, M.~Horibe, and A.~Hayashi,  
Phys. Rev. A {\bf 80}, 052322, (2009). 

\bibitem{Inui90}
T.~Inui,~Y. Tanabe, and Y.~Onodera, 
{\it Group Theory and Its Applications in Physics}
(Springer-Verlag, Heidelberg, 1990). 

\bibitem{Bennett92}
C.~H.~Bennett and S.~J.~Wiesner, 
Phys. Rev. Lett. {\bf 69}, 2881 (1992). 


\end{thebibliography}
\end{document}